\documentclass[aps,prb,twocolumn,superscriptaddress,amsmath,showpacs,tightenlines,longbibliography]{revtex4-1}

\usepackage{dcolumn}
\usepackage{graphicx}
\usepackage{mathrsfs}
\usepackage{subfigure}
\usepackage{booktabs}
\usepackage{amsmath}
\usepackage{dsfont}
\usepackage{amstext}
\usepackage{amssymb}
\usepackage{amsbsy}
\usepackage{bbm}
\usepackage{amsthm}
\usepackage{graphicx}
\usepackage{color}
\usepackage[colorlinks,citecolor=blue]{hyperref}

\usepackage{url}
\usepackage[colorlinks]{hyperref}
\hypersetup{%
	plainpages=true,
	breaklinks=true,  %not default in dvips mode, so we must specify
	hypertexnames=false,  %not ideal, but needed when pagenums duplicate (`i' vs. `1')
	pageanchor=true,
	colorlinks=true,
	linkcolor={blue},
	citecolor={red},
	urlcolor={blue},
	anchorcolor={black}
}

\bibpunct{[}{]}{,}{n}{}{}   % for prb/pra

 \makeatletter

\newcommand{\Rmnum}[1]{\expandafter\@slowromancap\romannumeral #1@}
\makeatother

\usepackage{CJKutf8}

\begin{document}
\begin{CJK*}{UTF8}{gbsn}

\title{Majorana corner states in a two-dimensional magnetic topological insulator  \\ on a high-temperature superconductor}
\author{Tao Liu}
\affiliation{Theoretical Quantum Physics Laboratory, RIKEN Cluster for Pioneering Research, Wako-shi, Saitama 351-0198, Japan}

\author{James Jun He}
\affiliation{RIKEN Center for Emergent Matter Science (CEMS), Wako, Saitama 351-0198, Japan}

\author{Franco Nori (野理)}
\email[E-mail: ]{fnori@riken.jp}
\affiliation{Theoretical Quantum Physics Laboratory, RIKEN Cluster for Pioneering Research, Wako-shi, Saitama 351-0198, Japan}
\affiliation{Department of Physics, University of Michigan, Ann Arbor, Michigan 48109-1040, USA}

%\date{{\small \today}}

%---------------------------------------------------------------------------

\begin{abstract}
Conventional $n$-dimensional topological superconductors (TSCs) have protected gapless $(n - 1)$-dimensional boundary states. In contrast to this, second-order TSCs are characterized by topologically protected gapless $(n - 2)$-dimensional states with usual gapped $(n - 1)$-boundaries. Here, we study a second-order TSC with a two-dimensional (2D) magnetic topological insulator (TI) proximity-coupled to a high-temperature superconductor, where Majorana bound states (MBSs) are localized at the corners of a square sample with gapped edge modes. Due to the mirror symmetry of the hybrid system considered here, there are two MBSs at each corner for both cases: $d$-wave and $s_{\pm}$-wave superconducting pairing. We present the corresponding topological phase diagrams related to the role of the magnetic exchange interaction and the pairing amplitude. A detailed analysis, based on edge theory, reveals the origin of the existence of MBSs at the corners of the 2D sample, which results from the sign change of the Dirac mass emerging at the intersection of any two adjacent edges due to pairing symmetry. Possible experimental realizations are discussed. Our proposal offers a promising platform for realizing MBSs and performing possible non-Abelian braiding in 2D systems.
\end{abstract}

\maketitle

\section{Introduction}
The study of nontrivial topological bands have led to the advent of a plethora of novel phases of matter characterized by topological invariants, which are independent of their microscopic details. These phases are characterized by a finite energy gap in the bulk and \mbox{protected} gapless states at their edges, with \mbox{unusual} properties. Recent years have seen a great deal of theoretical and experimental efforts towards the realization and exploration of Majorana zero modes (MZMs) in topological phases of quantum matter \cite{RevModPhys.82.3045, RevModPhys.83.1057, RPPJason2012, Beenakker2013, RPPMasatoshi2017, Ramon2017}. \mbox{MZMs} are zero-energy bound quasiparticles emerging at the boundaries of topological superconductors (TSCs), which are expected to exhibit exotic non-Abelian anyon \mbox{statistics}. This distinct feature makes MZMs promising for studying fault-tolerant topological quantum computations \cite{RevModPhys2008.80.1083, Alicea2011, Sarma2015}. \mbox{Several} promising condensed matter systems potentially hosting MZMs have been proposed, including: spin-orbit coupling semiconductor nanowire/superconductor hybrid structures \cite{PhysRevLett.105.177002, PhysRevLett.105.077001, Mourik1003, Deng1557, Zhang2018, Gul2018}, ferromagnetic atomic chains on superconductors \cite{PhysRevLett.111.147202,PhysRevLett.111.186805, Nadj-Perge602, Pawlak2016},  topological insulator/superconductor hybrid structures \cite{PhysRevLett.100.096407, PhysRevLett.104.040502, PhysRevB2015.92.075432, PhysRevB2016.94.125428, He294}, hybrid systems with unconventional superconductivity  \cite{PhysRevLett.103.107002, PhysRevLett.104.067001, PhysRevLett.111.056402, PhysRevB.97.064501, PhysRevLett.120.017001}, among others.

The nontrivial topological band structure of superconductor systems is the essential ingredient for the creation of MZMs in previous proposals, which is characterized by the bulk-boundary correspondence.  Very recently, the concept of higher-order topological insulators (TIs) \cite{PhysRevLett.110.046404, Benalcazar61,PhysRevB.96.245115, PhysRevLett.119.246401, PhysRevLett.119.246402,arXiv:1708.03636, PhysRevLett.120.026801, arXiv:1801.10050, arXiv:1801.10053, Imhof2018, arXiv:1802.02585,Franca2018} was put forward, where the usual \mbox{form} of the bulk-boundary correspondence is no longer applicable. As a new type of topological phase, it has no gapless surface states on three-dimensional (3D) insulators and gapless edge states on 2D ones. Nevertheless, the $n$-dimensional systems have protected gapless $(n - 2)$-dimensional states with the usual gapped $(n - 1)$-dimensional boundaries. For example, a second-order TI in 3D hosts 1D gapless modes in its hinges, while a second-order 2D TI has zero-energy states localized at its corners.

In terms of second-order TSCs in 2D, the MZMs will \mbox{emerge} at its \mbox{corners} i.e., Majorana corner states \mbox{(MCSs)}, which are localized at the intersection of two gapped topologically distinct edges. The study of MCSs is still at a very exploratory stage, and a few works are \mbox{recently} reported: high-temperature Majorana \mbox{Kramers} pairs with time-reversal symmetry localized at corners   \cite{arXiv:1803.08545, arXiv:1804.04711, Hsu2018}, \mbox{MCSs} in a $p$-wave superconductor with an in-plane external magnetic field \cite{PRBXYZhu2018}, Majorana bound states (MBSs) in a second-order Kitaev spin liquid \cite{arXiv:1803.08922}, as well as 2D and 3D second-order TSCs with $(p + i p)$ and $(p + i d)$ superconductors \cite{arXiv:1804.01531}.

In this article, we study a new kind of hybrid superconducting structure with a 2D magnetic TI and a high-temperature superconductor. This 2D magnetic TI shows a quantum anomalous Hall effect  and has been intensively investigated \cite{PhysRevLett.101.146802, PhysRevLett.111.136801, Yu61}. These can now be experimentally realized by introducing magnetic doping with Cr, V, or Mn ions \cite{Lee1316, Chang167, Mogieaao1669, Fan2014, Checkelsky2012, Chang2015, QAH_Mingda_Li2016}, or inducing proximity-induced ferromagnetism with a ferromagnetic insulator (FI) (i.e., TI/FI heterostructure) \cite{PhysRevLett.110.186807, QAH_Mingda_Li2016} to TI. Moreover, chiral MZMs are currently experimentally observed in a magnetic TI through the proximity effect to a conventional $s$-wave superconductor \cite{He294}. Additionally, the cuprate-based \cite{Zareapour2012, PhysRevB.91.235143, Wang2013} and iron-based  \cite{RevModPhys.83.1589, PhysRevLett.117.047001, arXiv:1706.06074, RepProgPhysFeSP2011, Zhangeaan4596} high-temperature superconductors have been experimentally reported to induce topological superconductivity. One important open question is whether a 2D magnetic TI \mbox{approximated} by a high-temperature superconductor can exhibit a second-order TSC hosting MBSs localized at their corners, and how the magnetic exchange interaction in the 2D magnetic TI influences the second-order topological features. Here we show that a second-order TSC can be achieved by a 2D magnetic TI grown on a cuprate-based or iron-based high-temperature superconductor, respectively. Although the hybrid superconductor system is in the topologically trivial regime with an insulating gap, there are MBSs localized at each corner of a square \mbox{sample}. The existence of Majorana corner states (MCSs) requires a magnetic insulator in a topologically nontrivial regime with protected chiral edges. These edge states can be gapped out once the high-temperature superconductor pairing (e.g., $d$-wave pairing) is introduced. Due to the superconducting pairing symmetry, the gapped two adjacent edges intersecting at corners have opposite Dirac mass, where MCSs are generated at one corner. Because the hybrid system considered here has mirror symmetry, there are two MBSs at each corner. In this article, in order to demonstrate this second-order TSC, we apply an intuitive edge argument. Moreover, we derive the topological phase diagrams involving the role of magnetic exchange interaction and pairing amplitude. The proposed second-order TSC provides an alternate new method for realizing MBSs. This suggests a promising platform for braiding MZMs in 2D systems, which may not be achieved for chiral Majorana modes in conventional 2D TSCs \cite{PhysRevB.97.104504}.

This paper is organized as follows: in Sec. $\textrm{\Rmnum{2}}$, we consider a minimal model on a square lattice describing a magnetic TI approximated by a high-temperature superconductor. \mbox{Sec. $\textrm{\Rmnum{3}}$} presents results of magnetic TIs grown on either a $d$-wave or an $s_{\pm}$-wave superconductor, respectively. An intuitive edge argument is given, and their topological phase diagrams are provided. Sec. $\textrm{\Rmnum{4}}$ describes their experimental feasibility, and concludes this paper.

\section{Model}
We here consider a minimal model on a square lattice, which describes a magnetic TI approximating a high-temperature superconductor, as shown in Fig.~1.  The tight-binding Hamiltonian is given by $H={{H}_{\textrm{t}}}+{{H}_{\textrm{so}}}+{{H}_{\textrm{z}}}+{{H}_{\textrm{sc}}}+\textrm{H.c.}$
\begin{align}
& {{H}_{\textrm{t}}} = {m}_{0} \sum\limits_{j,s}{c_{j,a,s}^{\dagger }}\sigma _{z}^{ab}{{c}_{j,b,s}} + {m}_{x} \sum\limits_{j,s}{c_{j,a,s}^{\dagger }}\sigma _{z}^{ab}{{c}_{j+x,b,s}} \nonumber \\
& ~~~~~~~ + {{m}_{y}}\sum\limits_{j,s}{c_{j,a,s}^{\dagger }}\sigma _{z}^{ab}{{c}_{j+y,b,s}} - \mu \sum\limits_{j,\sigma ,s}{c_{j,\sigma ,s}^{\dagger }}{{c}_{j,\sigma ,s}} , 
\end{align}
\begin{align}
& {{H}_{\textrm{so}}}  = -i{\lambda}_{\textrm{so}}\sum\limits_{j,v =x,y}{c_{j,a,\alpha }^{\dagger }}s_{v}^{\alpha \beta }\sigma _{x}^{ab}{{c}_{j+v,b,\beta }}  , 
\end{align}
\begin{align}
& {{H}_{\textrm{z}}} = {\lambda}_{z}\sum\limits_{j,\sigma }{\left(c_{j,\sigma ,\uparrow }^{\dagger}{{c}_{j,\sigma ,\uparrow }}-c_{j,\sigma ,\downarrow }^{\dagger }{{c}_{j,\sigma ,\downarrow }} \right)} , 
\end{align}
\begin{align}
& {{H}_{\textrm{sc}}} = \Delta_{0}\sum\limits_{j}{c_{j,\sigma ,\uparrow }^{\dagger }c_{j,\sigma ,\downarrow }^{\dagger }}+\frac{{\Delta}_{x}}{2}\sum\limits_{j}{c_{j,\sigma ,\uparrow }^{\dagger }c_{j \pm x,\sigma ,\downarrow }^{\dagger}} \nonumber \\
& ~~~~~~~ +\frac{{\Delta}_{y}}{2}\sum\limits_{j}{c_{j,\sigma ,\uparrow}^{\dagger }c_{j \pm y,\sigma ,\downarrow }^{\dagger }} .
\end{align}
where these therms are the kinetic term $H_\textrm{t}$, spin-orbit coupling $H_{\textrm{so}}$, Zeeman coupling $H_\textrm{z}$, and superconducting pairing $H_{\textrm{sc}}$, respectively. Also, $s_i$ and $\sigma_i$ are Pauli matrices denoting the electron spins $(\uparrow, \downarrow)$ and orbitals $(a,b)$, respectively. The $c_{j,a ,s}$ is the fermion operator at site $j$, $m_0$ is the orbital-dependent on-site energy, and $m_x$ and $m_y$ are the intra-orbital hopping amplitudes along $\textit{x}$ and $\textit{y}$ axis, while $\mu$ is the chemical potential. Moreover, $\lambda_{\textrm{so}}$ is the spin-orbital coupling strength; and $\lambda_z$ is the exchange field amplitude along the $\textit{z}$ axis induced by the magnetization. The superconducting pairing terms $\Delta_0$, $\Delta_x$, and $\Delta_y$ are combined to characterize $d_{x^2-y^2}$ and $s_{\pm}$ wave pairing.

According to \mbox{Eqs.~(1)--(4)}, the \mbox{Bogoliubov-de} Gennes (BdG) Hamiltonian can be written as $\mathcal{H}_{\textrm{BdG}} = \sum_{\mathbf{k}} \Psi_{\mathbf{k}}^\dagger H_{\textrm{BdG}}(\mathbf{k}) \Psi_{\mathbf{k}}$, where $\Psi_{\mathbf{k}} = (c_{\mathbf{k}, a,\uparrow}, c_{\mathbf{k}, a,\downarrow}, c_{\mathbf{k}, b,\uparrow}, c_{ \mathbf{k}, b,\downarrow}, c_{\mathbf{\textrm{-}k}, a,\downarrow}^{\dagger}, \textrm{-}c_{\mathbf{\textrm{-}k}, a,\uparrow}^{\dagger}, c_{ \mathbf{\textrm{-}k}, b,\downarrow}^{\dagger}, \textrm{-}c_{\mathbf{\textrm{-}k}, b,\uparrow}^{\dagger})^T$,
\begin{align}
H_{\textrm{BdG}}(\mathbf{k}) & =   m(k) \sigma_{z} \tau_z + \lambda_{\textrm{so}} \left[\sin(k_x) s_x + \sin(k_y) s_y \right] \sigma_{x} \tau_z \nonumber \\
& ~~~  + \lambda_{z} s_z - \mu \tau_z + \Delta(k) \tau_x ,
\end{align}
here $m(k)$ and $\Delta(k)$ are
\begin{align}
& m(k) = m_0 + m_x \cos(k_x) + m_y \cos(k_y) , \\
& \Delta(k) = \Delta_0 + \Delta_x \cos(k_x) + \Delta_y \cos(k_y) ,
 \end{align}
where $\tau_i$ ($i = x, y, z$) are Pauli matrices in the Nambu particle-hole space.

The Hamiltonian $H_{\textrm{BdG}}(\mathbf{k})$ is invariant under a particle-hole symmetry $\Theta = \tau_y s_y \mathcal{K}$, with $\mathcal{K}$ being the complex conjugation operator,  a mirror-reflection symmetry $\mathcal{M}_z = i s_z \sigma_z$, a four-fold rotational symmetry $C_4 = \textrm{e}^{-i \frac{\pi}{4}s_z}$ and an inversion symmetry $\mathcal{P} = \sigma_z$
\begin{align}
& \Theta H_{\textrm{BdG}}(k_x, k_y) \Theta^{-1} = -H_{\textrm{BdG}}(-k_x, -k_y) ,\\
& \mathcal{M}_z H_{\textrm{BdG}}(k_x, k_y) \mathcal{M}_z^{-1} = H_{\textrm{BdG}}(k_x, k_y) ,\\
& C_4 H_{\textrm{BdG}}(k_x, k_y) C_4^{-1} = H_{\textrm{BdG}}(-k_y, k_x) ,\\
& \mathcal{P} H_{\textrm{BdG}}(k_x, k_y) \mathcal{P}^{-1} = H_{\textrm{BdG}}(-k_x, -k_y) .
\end{align}
\begin{figure}
	\centering
	\includegraphics[width=8.4cm]{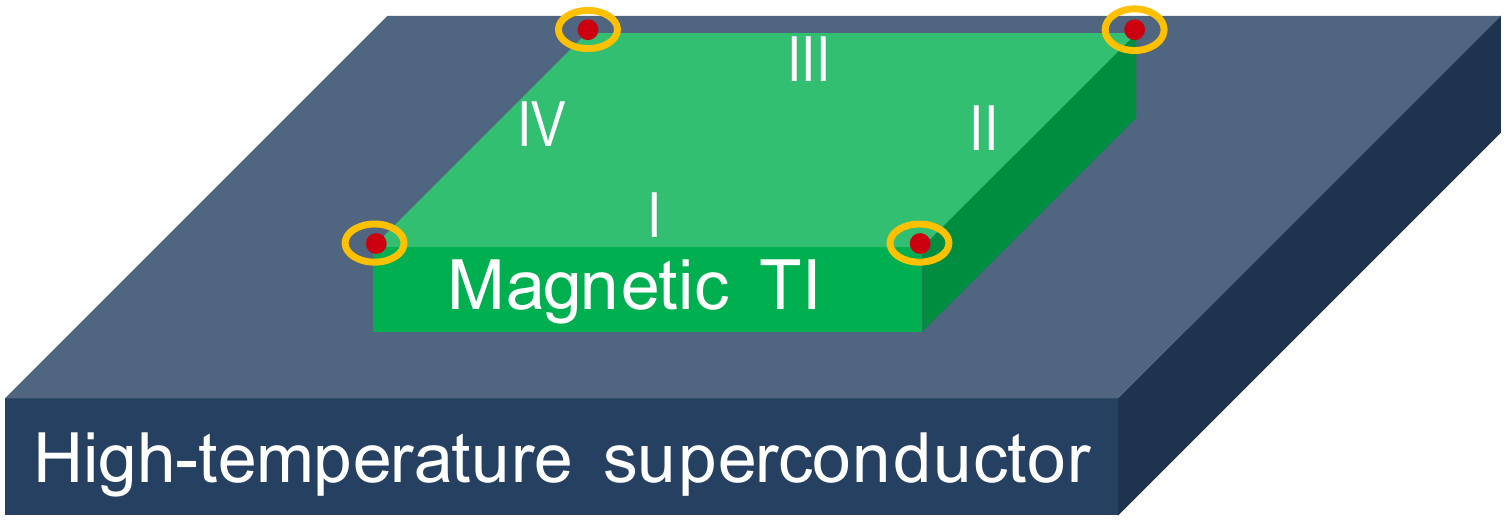}
	\caption{Schematic view showing a magnetic TI approximated by a high-temperature superconductor. The Majorana bound states (MBSs) are located at the four corners of the green square sample. $\textrm{\Rmnum{1}}$, $\textrm{\Rmnum{2}}$, $\textrm{\Rmnum{3}}$ and $\textrm{\Rmnum{4}}$ label the four edges of the lattice.}\label{fig1}
\end{figure}

\section{Results}
\subsection{$d$-wave pairing}
We first consider a magnetic TI grown on a $d$-wave cuprate high-temperature superconductor that has been widely investigated in experiments \cite{Zareapour2012, PhysRevB.91.235143, Wang2013}. For a $d$-wave superconductor with $d_{x^2-y^2}$ wave symmetry, the pairing amplitude satisfies
\begin{align}
\Delta_{0} = 0, ~~~\Delta_{x} = - \Delta_{y} = \Delta_{1} .
\end{align}

To explore whether the hybrid system of magnetic TI/$d$-wave superconductor exhibits second-order non-trivial topological phases, which support Majorana bound states at each corner of a square sample,  we first calculate the energy-band spectrum of the system. The 2D magnetic insulator is in the topologically non-trivial regime when the system parameters satisfy 
\begin{align}
\vert \vert m_x \vert  - \vert m_y \vert \vert < \vert m_0 ~\pm~ \lambda_z \vert < \vert m_x \vert + \vert m_y \vert .
\end{align}
Figure 2(a) and 2(b) show the energy-band structure of a 2D magnetic TI nanoribbon along the  $x$ and  $y$ directions, respectively. The red lines represent two degenerate gapless chiral edge states characterized by the Chern number $\mathcal{N} = 2$. The zero-energy edge states at the $y$ and $x$ directions exist at the $k_x =0$ and $k_y =0$ points with the parameters considered, respectively.

When a $d_{x^2-y^2}$ pairing is added to the magnetic TI, the chiral edges are gapped out (see red lines in Fig.~2(c) and 2(d)). In this case, the hybrid system is in a topologically trivial regime with  $\mathcal{N} = 0$. However, by calculating the eigen-energies of a finite square sample, two quite localized zero-energy states emerge at each corner, as shown in Fig.~2(e). Due to particle-hole symmetry $\Theta$, these zero-energy corner states are MZMs known as Majorana corner states (MCSs). The inset figure in Fig.~2(e) exibits the symmetrical eigen-energies with particle and hole bands. Figure 3 shows the BdG energy spectrum with open boundary conditions in the $x$ and $y$ directions as a function of $\lambda_z$. The eight-fold degenerate localized zero-energy MCSs states are indicated by the red curves, which exist only in a finite amplitudes of the exchange field.

\begin{figure}[!tb]
\centering
\includegraphics[width=8.4cm]{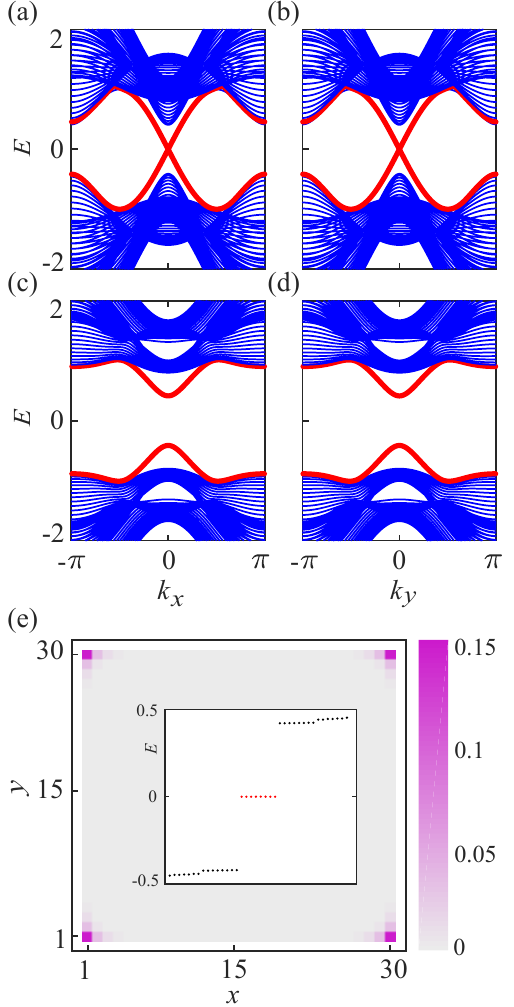}
\caption{Energy-band structure of a 2D magnetic TI nanoribbon along the (a) $x$ and (b) $y$ directions, respectively. The red curves in (a) and (b) denote two degenerate gapless chiral edge states. The zero-energy edge states exist at the $k_x = 0$ and $k_y = 0$ points, respectively. The BdG spectrum with a $d_{x^2-y^2}$ wave pairing along the (c) $x$  and (d) $y$  directions, respectively. In the presence of $\textit{d}$-wave pairing, the edge states are gapped out (red curves). (e) The probability density distributions of the BdG wavefunctions with zero energies for a sample size with $30 \times 30$. There are two MBSs localized at each corner due to the mirror symmetry of the BdG Hamiltonian. The inset shows the eigen-energies of the same sample with energies around zero. Note that there are eight zero energy modes in the gap shown as eight red dots. The parameters are chosen as $m_0 = -0.8$, $\lambda_{z} = 0.4$, $\lambda_{\textrm{so}}=1$, $m_x=m_y=1$, $\mu = 0$, and $\Delta_{1} = 0.5$.}\label{fig2}
\end{figure}
\begin{figure}[!tb]
	\centering
	\includegraphics[width=8.4cm]{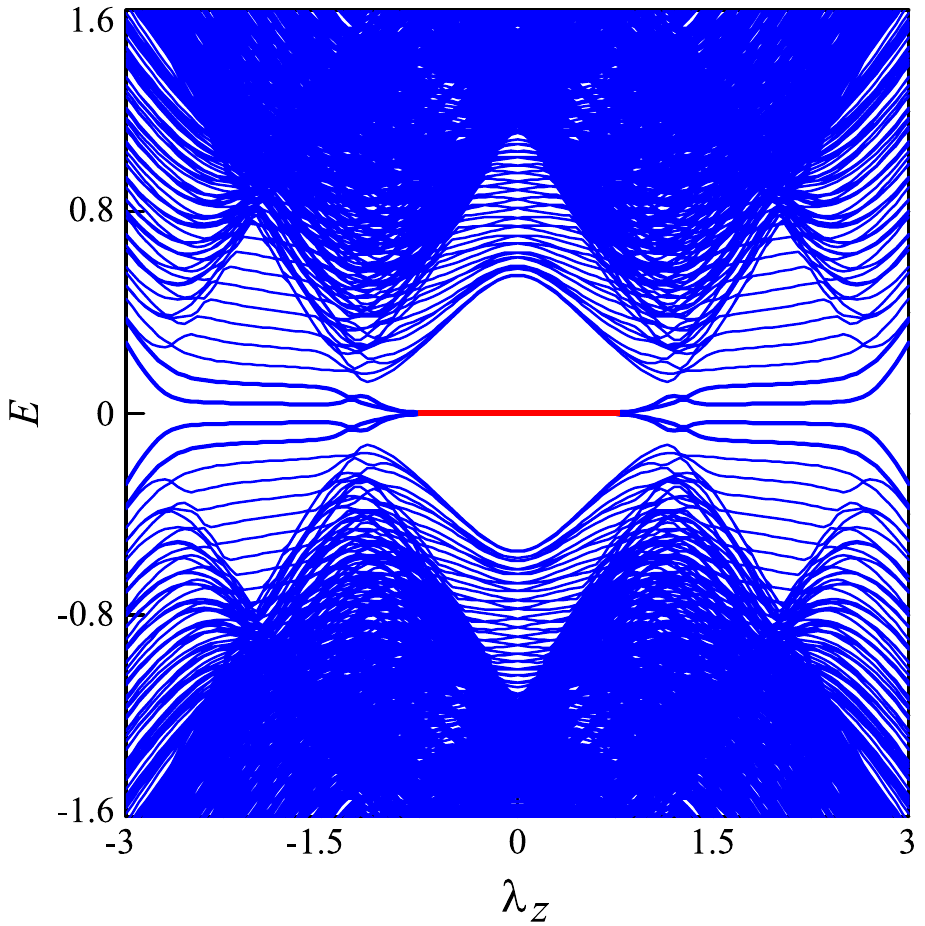}
	\caption{BdG energy spectrum with open boundary conditions in the $x$ and $y$ directions as a function of $\lambda_z$. The sample size is $20 \times 20$. The red line corresponds to eight-fold degenerate MCSs, which exist only for finite amplitudes of the exchange field. The parameters are chosen as $m_0 = -0.8$, $\lambda_{\textrm{so}}=1$, $m_x=m_y=1$, $\mu = 0$, and $\Delta_{1} = 0.5$.}\label{fig3}
\end{figure}

Figure 4 shows the topological phase diagram of the magnetic TI/$d_{x^2-y^2}$ high-temperature superconductor hybrid system in the $(m_0, \lambda_z)$ plane, which reveals three distinct phases: (i) Second-order topological superconductor with MCSs, (ii) Chiral MZMs characterized by a finite Chern number $\mathcal{N}$, and (iii) Topologically trivial states $\mathcal{N} = 0$ with zero chiral MZMs and MCSs. The phase boundaries are determined by the zero eigen-energy of the BdG Hamiltonian in Eq.~(5) at the four corners of the Brillouin zone of a square lattice i.e. $\varGamma = (0, 0)$, $X = (0, \pi)$, $Y = (\pi, 0)$, and $M = (\pi, \pi)$, where $\varGamma$ and $M$ are two high-symmetry points. The energies at these points are
\begin{align}
E_{\varGamma}  = \pm\mu ~\pm~ (m_0 + m_x + m_y) ~\pm~ \lambda_{z}  ,
\end{align}
\begin{align}
E_{M}  = \pm\mu ~\pm~(m_0 - m_x - m_y) ~\pm~ \lambda_{z} ,
\end{align}
\begin{align}
E_{X}  = \pm\sqrt{(\mu ~\pm~ m_0 ~\pm~ m_x ~\mp~ m_y)^2 + 4 \Delta_{1}^2)} ~\pm~ \lambda_{z}  ,
\end{align}
\begin{align}
E_{Y}  = \pm\sqrt{(\mu ~\pm~ m_0 ~\mp~ m_x ~\pm~ m_y)^2 + 4 \Delta_{1}^2)} ~\pm~ \lambda_{z} .
\end{align}
In addition, the existence of MCSs requires the 2D magnetic insulator in the topologically non-trivial regime [see Eq.~(13)]. All these determine the topological phase diagram shown in Fig.~4.

\begin{figure}[!tb]
	\centering
	\includegraphics[width=8.4cm]{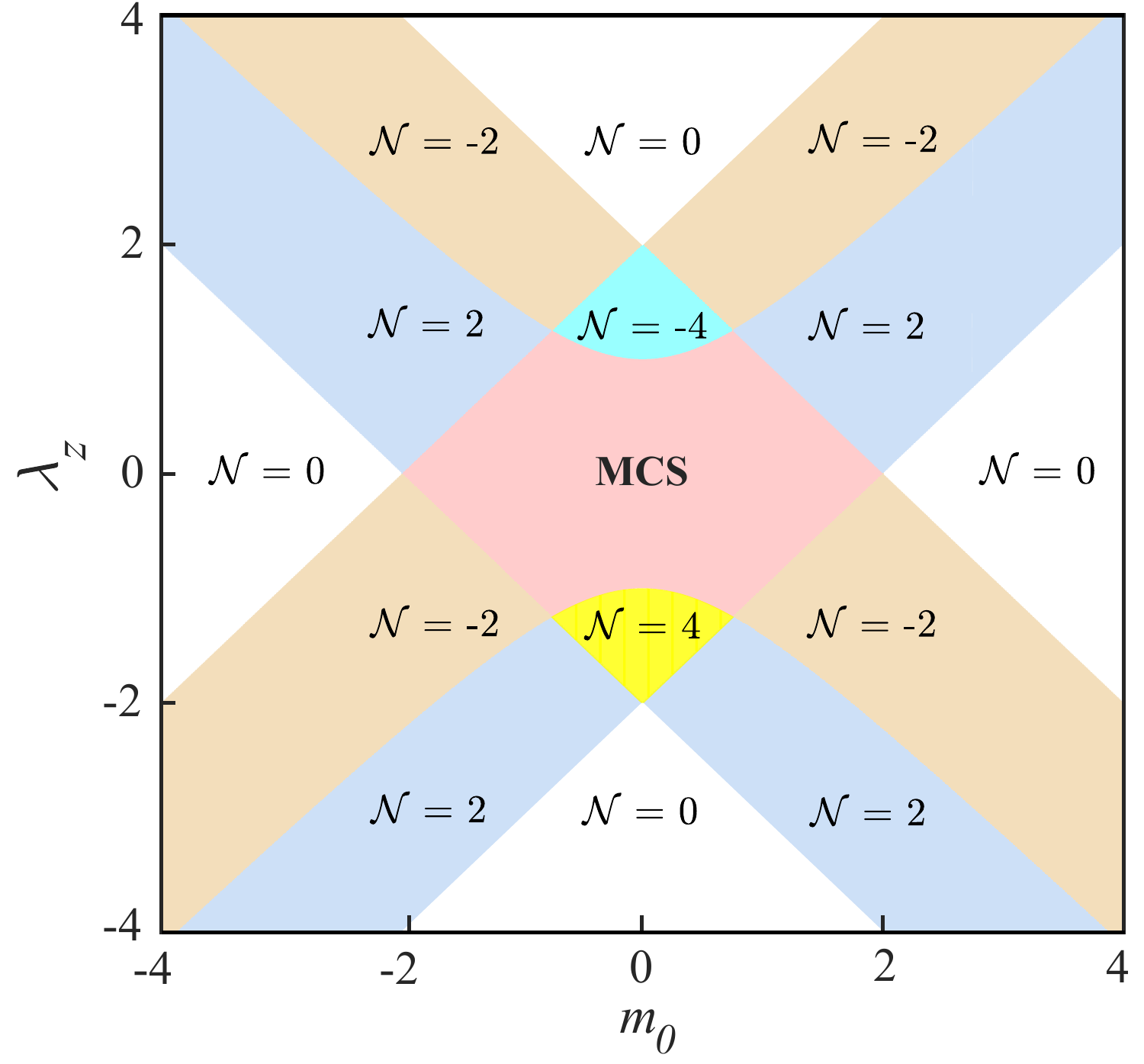}
	\caption{Topological phase diagram of the magnetic TI/$d_{x^2-y^2}$ high-temperature superconductor hybrid system in the $(m_0, \lambda_z)$ plane, which reveals three distinct phases: (i) Second-order topological superconductor with MCSs, (ii) Chiral MZMs characterized by a finite Chern number $\mathcal{N}$, and (iii) Topologically trivial states $\mathcal{N} = 0$ with zero chiral MZMs and MCSs. The parameters are chosen as $\lambda_{\textrm{so}}=1$, $m_x=m_y=1$, $\mu = 0$, and $\Delta_{1} = 0.5$.}\label{fig4}
\end{figure}

In order to intuitively understand the appearances of MBSs at the corners, we apply the edge theory (see e.g.,\cite{RevModPhys.83.1057, PRBXYZhu2018, arXiv:1803.08545}). Due to the mirror-reflection symmetry $\mathcal{M}_z$, the BdG Hamiltonian can be written in the block-diagonal form by a unitary transformation $U$
\begin{align}
U H_{\textrm{BdG}} U^{-1} = \left(\begin{matrix}
H_{+}(\mathbf{k}) & 0 \\
0 & H_{-}(\mathbf{k}) \\  \end{matrix}\right) ,
\end{align}
where $H_{+}(\mathbf{k})$ acts on the $+i$ mirror subspace, $H_{-}(\mathbf{k})$ acts on the $-i$ mirror subspace, and they are expressed as
\begin{align}
H_{\pm}(\mathbf{k}) & = m(k) \eta_{z} \tau_z ~\pm~ \lambda_{z} \eta_{z} - \mu \tau_z + \Delta(k) \tau_x \nonumber \\
& ~~~ + \lambda_{\textrm{so}} \left[\sin(k_x) \eta_x ~\pm~ \sin(k_y) \eta_y \right] \tau_z ,
\end{align}
and $\eta_i$ are the Pauli matrices. Note that the similar mirror symmetry, in combination with time-reversal symmetry, was taken into account in topological crystalline superconductors \cite{PhysRevLett.111.056403}.

In order to solve the effective Hamiltonian of the edge states, we consider the continuum model of the lattice Hamiltonian by expanding its wave vector $\mathbf{k}$ in Eqs.~(18)--(19) to second order around the $\varGamma = (0, 0)$ point (we can also expand $\mathbf{k}$ around the other high-symmetry points, e.g., $M$, which will not influence the discussions below).
\begin{align}
H_{\textrm{c}}(\mathbf{k}) = \left(\begin{matrix}
H_{+}^c(\mathbf{k}) & 0 \\
0 & H_{-}^c(\mathbf{k}) \\  \end{matrix}\right) ,
\end{align}
where
\begin{align}
H_{\pm}^c(\mathbf{k}) & =\left[m_1 - \dfrac{1}{2} (m_x k_{x}^2 + m_y k_{y}^2)\right] \eta_{z} \tau_z ~\pm~ \lambda_{z} \eta_{z} - \mu \tau_z    \nonumber \\
& ~~~  + \lambda_{\textrm{so}} \left[k_x \eta_x ~\pm~ k_y \eta_y \right] \tau_z - \dfrac{\Delta_1}{2}  (k_{x}^2 - k_{y}^2) \tau_x ,
\end{align}
where $m_1 = m_0 + m_x + m_y$.

We first solve the edge $\textrm{\Rmnum{1}}$ of the four edges (see Fig.~1). By expressing $k_y$ as $-i \partial_y$, and treating the pairing terms as perturbation (which is valid when the pairing amplitude is relatively small), we can rewrite the Hamiltonian $H_{\pm}^c$ as  $H_{\pm}^c = H_{\pm}^1 + H_{\pm}^2 $:
\begin{align}
& H_{\pm}^1 (k_x, -i \partial_y)  = \left(m_1 + m_y {\partial^2_y}/2\right) \eta_{z} \tau_z ~\pm~ \lambda_{z} \eta_{z} - \mu \tau_z    \nonumber \\
& ~~~~~~~~~~~~~~~~~~~~~   ~\mp~ i \lambda_{\textrm{so}} \eta_y \tau_z {\partial_y}  , \\
& H_{\pm}^2 (k_x,-i \partial_y)  = \lambda_{\textrm{so}} k_x \eta_x \tau_z - (\Delta_1/2) \tau_x {\partial^2_y}  ,
\end{align}
where we have already neglected the insignificant $k_{x}^2$ terms.
To obtain the eigenvalue equation $H_{\pm}^1 \phi_{\pm}(y) = E_{\pm} \phi_{\pm}(y)$, with $E_{\pm} = 0$ under boundary conditions $\phi_{\pm}(0) = \phi_{\pm}(+\infty) =0$, we write the solution in the following form
\begin{align}
\phi_{\pm}(y) = \mathcal{N}_y\sin(\alpha y) \textrm{e}^{-\beta y} \textrm{e}^{i k_x x} \chi_{\pm} ,
\end{align}
where the normalization constant $\mathcal{N}_y = 2 \sqrt{\beta (\alpha^2 + \beta^2)/\alpha^2}$. The eigenvector $\chi_{\pm}$ satisfies $\eta_{x} \chi_{\pm} = \mp \textrm{sgn} (m_y) \chi_{\pm}$. For the sake of simplicity, we assume $\lambda_{\textrm{so}} > 0$ in our discussions unless otherwise specified. Then the effective Hamiltonian for the edge $\textrm{\Rmnum{1}}$ can be obtained in this basis as
\begin{align}
\mathcal{H}_{\pm}^{\textrm{\Rmnum{1}}}  = \int_{0}^{+\infty} \!\! \phi_{\pm}^*(y) H_{\pm}^2 \phi_{\pm}(y) \: dy .
\end{align}
Therefore, we have
\begin{align}
& \mathcal{H}_{\pm}^{\textrm{\Rmnum{1}}} = \mp~ \textrm{sgn} (m_y) \lambda_{\textrm{so}} k_x \tau_z + \frac{\Delta_{1}}{2} \left(\alpha_1^2 + \beta_1^2 \right) \tau_x ,
\end{align}
where $\alpha_1^2 + \beta_1^2 = 2(m_1 ~\pm~ \lambda_z ~\pm~ \mu)/m_y$.

The effective Hamiltonian for the edges $\textrm{\Rmnum{2}}$, $\textrm{\Rmnum{3}}$ and $\textrm{\Rmnum{4}}$ can be obtained by the same procedures:
\begin{align}
& \mathcal{H}_{\pm}^{\textrm{\Rmnum{2}}} = \mp~ \textrm{sgn} (m_x) \lambda_{\textrm{so}} k_y \tau_z - \frac{\Delta_{1}}{2} \left(\alpha_2^2 + \beta_2^2 \right) \tau_x , \\
& \mathcal{H}_{\pm}^{\textrm{\Rmnum{3}}} = \pm~ \textrm{sgn} (m_y) \lambda_{\textrm{so}} k_x \tau_z + \frac{\Delta_{1}}{2} \left(\alpha_1^2 + \beta_1^2 \right) \tau_x , \\
& \mathcal{H}_{\pm}^{\textrm{\Rmnum{4}}} = \pm~ \textrm{sgn} (m_x) \lambda_{\textrm{so}} k_y \tau_z - \frac{\Delta_{1}}{2} \left(\alpha_2^2 + \beta_2^2 \right) \tau_x ,
\end{align}
where $\alpha_2^2 + \beta_2^2 = 2(m_1 ~\pm~ \lambda_z ~\pm~ \mu)/m_x$.

The first kinetic terms of the effective Hamiltonian in Eqs.~(26)--(29) describe the gapless edge states, which are gapped out by the second terms with Dirac mass. Moreover, at the mirror subspace with the Hamiltonians $H_{+}^i$ and $H_{-}^i$ ($i = \textrm{\Rmnum{1}} \sim \textrm{\Rmnum{4}}$), due to mirror-reflection symmetry, the Dirac mass terms change sign along four edges, edges $\textrm{\Rmnum{1}}$ to $\textrm{\Rmnum{4}}$, resulting from the $d_{x^2-y^2}$ pairing symmetry; i.e., any two adjacent edge states have opposite Dirac mass, while there are the same signs for the first kinetic terms along the anticlockwise direction of the edges. As a result,  there is one MBS at each corner of a square sample within each mirror subspace of the Hamiltonian (see the Jackiw-Rebbi model \cite{PhysRevD.13.3398}); i.e., two MBSs at each corner. By comparing the coefficients of the Dirac mass terms, containing $\alpha_1^2 + \beta_1^2$ and $\alpha_2^2 + \beta_2^2$, for the two adjacent edge states in Eqs.~(26)--(29), in order to ensure the existence of Majorana corner states, $m_x$ and $m_y$ should satisfy the relation $m_x m_y > 0$. Note that the BdG system supports two MCSs at each corner in the whole regime of the second-order topological phase considered here due to the mirror-reflection symmetry $\mathcal{M}_z$ for the Hamiltonian $H_{\textrm{BdG}}(\mathbf{k})$. When this mirror-reflection symmetry is broken, a single MCS at each corner can be achieved (see Appendix).

\subsection{$s_{\pm}$ wave pairing}
Here we consider the magnetic TI approximated by an $s_{\pm}$ superconducting pairing, which is relevant for iron-based high-temperature superconductors \cite{RevModPhys.83.1589, PhysRevLett.117.047001, arXiv:1706.06074, RepProgPhysFeSP2011, Zhangeaan4596}. The pairing amplitude of an $s_{\pm}$ wave superconductor satisfies
\begin{align}
\Delta_{x} = \Delta_{y} = \Delta_{2} .
\end{align}

As in the case for a $d$-wave superconductor, we first consider the energy-band spectrum of the system. Figures 5(a) and 5(b) show the energy-band structure of a 2D magnetic TI nanoribbon along the $x$ and  $y$ directions, respectively. But, in contrast to the case for hybrid systems of magnetic TI/$d$-wave superconductors, the zero-energy edge states at the $y$ and $x$ directions exist at the $k_x = 0$ and $k_y = \pi$ points for magnetic TI/$s_{\pm}$-wave superconductor hybrid systems when $m_y$ is set to a negative value.  In the presence of the $s_{\pm}$ pairing, the edges are gapped out [see the red curves in Fig.~5(c) and 5(d)], where the hybrid system enters the topologically trivial regime. However, similarly to the case for $d$-wave superconductor, each corner of the finite-size sample supports two localized MBSs [see Fig. 5(e)]. Moreover, when $m_y$ is set to have the same sign as $m_x$, there are no MCSs. Note that the system may support a single MCS at each corner when the mirror-reflection symmetry $\mathcal{M}_z$ is broken (see Appendix).

\begin{figure}[!tb]
	\centering
	\includegraphics[width=8.2cm]{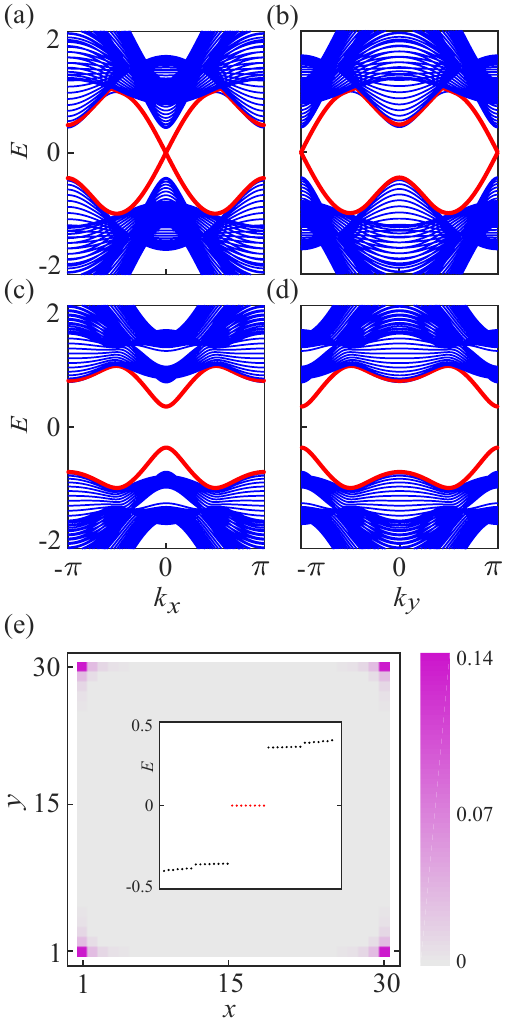}
	\caption{Energy-band structure of a 2D magnetic TI nanoribbon along the (a) $x$ and (b) $y$ directions, respectively. The red curves in (a) and (b) denote two degenerate gapless chiral edge states. The zero-energy edge states exist at the $k_x = 0$ and $k_y = \pi$ points, respectively. The BdG spectrum with an $s_{\pm}$ wave pairing along the (c) $x$  and (d) $y$  directions, respectively. In presence of $s_{\pm}$-wave pairing, the edge states are gapped out (red curves). (e) The probability density distributions of BdG wavefunctions with zero energies for a sample of size $30 \times 30$. There are two MBSs localized at each corner due to the mirror symmetry of the BdG Hamiltonian. The inset shows eigen-energies of the same sample with energies around zero. Note that there are eight zero energy modes in the gap shown as eight red dots. The parameters are chosen as $m_0 = -0.8$, $\lambda_{z} = 0.4$, $\lambda_{\textrm{so}}=1$, $m_x=-m_y=1$, $\mu = 0$, $\Delta_{0} = 0$, and $\Delta_{2} = 0.4$.}\label{fig5}
\end{figure}

In order to understand the existence of MCSs with $s_{\pm}$ wave pairing, we also consider the edge theory. In this part, we consider the continuum model of the lattice Hamiltonian by expanding its wave vector $\mathbf{k}$ in Eqs.~(18)--(19) to second order around the $X = (0, \pi)$ point of the Brillouin zone, obtaining
\begin{align}
H_{\textrm{c}}(\mathbf{k}) = \left(\begin{matrix}
H_{+}^c(\mathbf{k}) & 0 \\
0 & H_{-}^c(\mathbf{k}) \\  \end{matrix}\right) ,
\end{align}
where
\begin{align}
H_{\pm}^c(\mathbf{k}) & = \left[m_2 - \dfrac{1}{2} (m_x k_{x}^2 - m_y k_{y}^2)\right] \eta_{z} \tau_z ~\pm~ \lambda_{z} \eta_{z}     \nonumber \\
& ~~~  + \lambda_{\textrm{so}} \left[k_x \eta_x ~\pm~ k_y \eta_y \right] \tau_z - \mu \tau_z \nonumber \\ & ~~~ + \left[\Delta_{0} - \dfrac{\Delta_2}{2}  (k_{x}^2 - k_{y}^2)\right] \tau_x ,
\end{align}
and $m_2 = m_0 + m_x - m_y$.

Then, the effective Hamiltonian for the edges $\textrm{\Rmnum{1}}$, $\textrm{\Rmnum{2}}$, $\textrm{\Rmnum{3}}$ and $\textrm{\Rmnum{4}}$ can be obtained as
\begin{align}
& \mathcal{H}_{\pm}^{\textrm{\Rmnum{1}}} = \mp~ \textrm{sgn} (m_y) \lambda_{\textrm{so}} k_x \tau_z + \left[\Delta_{0}+\frac{\Delta_{2}}{2} (\alpha_3^2 + \beta_3^2)\right] \tau_x  , \\
& \mathcal{H}_{\pm}^{\textrm{\Rmnum{2}}} = \pm~ \textrm{sgn} (m_x) \lambda_{\textrm{so}} k_y \tau_z + \left[\Delta_{0}-\frac{\Delta_{2}}{2} (\alpha_4^2 + \beta_4^2)\right] \tau_x , \\
& \mathcal{H}_{\pm}^{\textrm{\Rmnum{3}}} = \pm~ \textrm{sgn} (m_y) \lambda_{\textrm{so}} k_x \tau_z + \left[\Delta_{0}+\frac{\Delta_{2}}{2} (\alpha_3^2 + \beta_3^2)\right] \tau_x  , \\
& \mathcal{H}_{\pm}^{\textrm{\Rmnum{4}}} = \mp~ \textrm{sgn} (m_x) \lambda_{\textrm{so}} k_y \tau_z + \left[\Delta_{0}-\frac{\Delta_{2}}{2} (\alpha_4^2 + \beta_4^2)\right] \tau_x , 
\end{align}
where $\alpha_3^2 + \beta_3^2 = -2(m_2 ~\pm~ \lambda_z ~\pm~ \mu)/m_y$, and $\alpha_4^2 + \beta_4^2 = 2(m_2 ~\pm~ \lambda_z ~\pm~ \mu)/m_x$.

First, according to Eqs.~(33)--(36), in contrast to the case of $d$-wave superconductors, the existence of Majorana corner states requires $m_x$ and $m_y$ to satisfy $m_x m_y < 0$ for hybrid systems with $s_{\pm}$ superconducting pairing.

Second, within each mirror subspace of the BdG Hamiltonian, in order to ensure the opposite Dirac mass terms for any two gapped adjacent edge states, the following criterion should be satisfied:
\begin{align}
\left[\Delta_{0} + \frac{\Delta_{2}}{2} \left(\alpha_3^2 + \beta_3^2 \right) \right] \left[\Delta_{0} - \frac{\Delta_{2}}{2} \left(\alpha_4^2 + \beta_4^2 \right) \right] < 0 .
\end{align}
Therefore, we have
\begin{align}
\left[\frac{\Delta_{0}}{\Delta_{2}} - \frac{m_2 ~\pm~ \lambda_z ~\pm~ \mu}{m_y}\right] \left[\frac{\Delta_{0}}{\Delta_{2}} - \frac{m_2 ~\pm~ \lambda_z ~\pm~ \mu}{m_x} \right] < 0 .
\end{align}

Equations (13) and (38) determine the system parameters, including the pairing amplitude and magnetic exchange interaction, required for the existence of MCSs for $s_{\pm}$ superconducting pairing. As an example, according to Eqs.~(13) and (38), the criterion of \mbox{$\Delta_{0}/\Delta_{2} < 1$} should be satisfied to ensure the existence of MCSs within each mirror subspace of the Hamiltonian, if $m_0 = -1$ and $m_x = -m_y = -1$. 

Figure 6(a) shows the emergence of MCSs by computing the probability density distribution of the BdG wavefunctions for the magnetic TI/$s_{\pm}$ superconductor hybrid system with the parameters $\lambda_z$, $\Delta_{0}$, $\Delta_{2}$, $m_x$, and $m_y$ satisfying all of these criteria. There are then two MBSs localized at each corner. A finite chemical potential $\mu$ within the limit of Eq.~(38) will not destroy the MCSs, as shown in Fig.~6(b).

\begin{figure}[!tb]
	\centering
	\includegraphics[width=8.4cm]{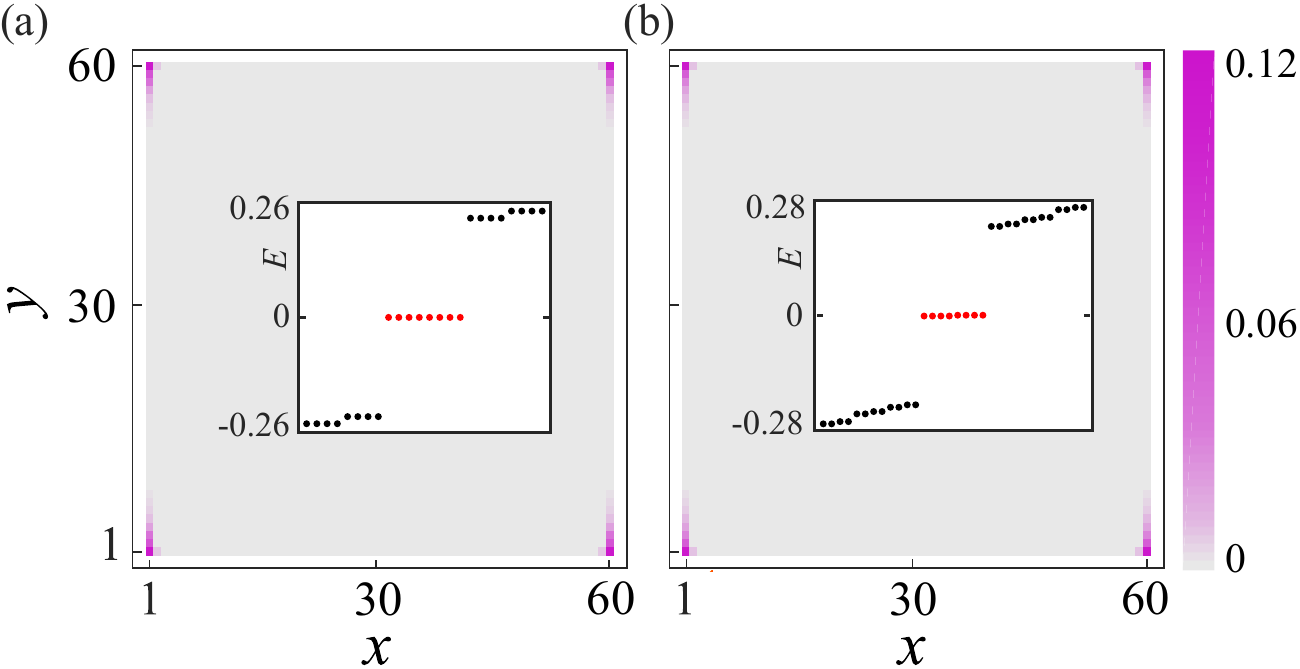}
	\caption{The probability density distributions (with scale shown in the right vertical bar) of the BdG wavefunctions with zero energies for a sample of size of $60 \times 60$, with different chemical potentials: (a) $\mu = 0$, and (b) $\mu = 0.1$. There are two MBSs localized at each corner. A finite chemical potential $\mu$ will not destroy the MCSs. The inset shows the eigen-energies of the same sample with the energies around zero. Note that there are eight zero energy modes in the gap shown as eight red dots. Other parameters are chosen as $m_0 = -0.8$, $\lambda_{z} = 0.1$, $\lambda_{\textrm{so}}=1$, $m_x=-m_y=1$, $\Delta_{0}/\Delta_{2} = 0.6$, and $\Delta_{2} = 0.4$.}\label{fig6}
\end{figure}

In terms of topological phase diagram, parts of the phase boundaries are determined by the zero eigen-energy of the BdG Hamiltonian in Eq.~(5) at the four corners of the Brillouin zone of a square lattice i.e., $\varGamma = (0, 0)$, $X = (0, \pi)$, $Y = (\pi, 0)$, and $M = (\pi, \pi)$. The energies for a TI/$s_{\pm}$ hybrid system at these points are

\begin{align}
& E_{\varGamma}  = \pm \sqrt{(\mu ~\mp~ m_0 ~\mp~ m_x ~\mp~ m_y)^2 + (\Delta_{0} + 2 \Delta_{2})^2}  \nonumber \\
& ~~~~~~~ ~\pm~ \lambda_z  , \\
& E_{M}  =   \pm \sqrt{(\mu ~\pm~ m_0 ~\mp~ m_x ~\mp~ m_y)^2 + (\Delta_{0} - 2 \Delta_{2})^2} \nonumber \\
& ~~~~~~~ ~\pm~ \lambda_z  , \\
& E_{X}  = \pm \sqrt{(\mu ~\pm~ m_0 ~\pm~ m_x ~\mp~ m_y)^2 + \Delta_{0}^2} ~\pm~ \lambda_z  ,  \\
& E_{Y}  = \pm \sqrt{(\mu ~\mp~ m_0 ~\pm~ m_x ~\mp~ m_y)^2 +\Delta_{0}^2} ~\pm~ \lambda_z .
\end{align}
In addition, the phase boundaries are simultaneously determined by Eq.~(13) and (38). Figure 7 shows the topological phase diagram of the magnetic TI/$s_{\pm}$ high-temperature superconductor hybrid system in the $(\lambda_z, \Delta_{0}/\Delta_{2})$ plane. As in the case of $d$-wave superconductors, there are three distinct phases: (i) Second-order topological superconductor with MCSs, (ii) Chiral MZMs characterized by a finite Chern number $\mathcal{N}$, and (iii) Topologically trivial states $\mathcal{N} = 0$ with zero chiral MZMs and MCSs. 

\begin{figure}[!tb]
	\centering
	\includegraphics[width=8.4cm]{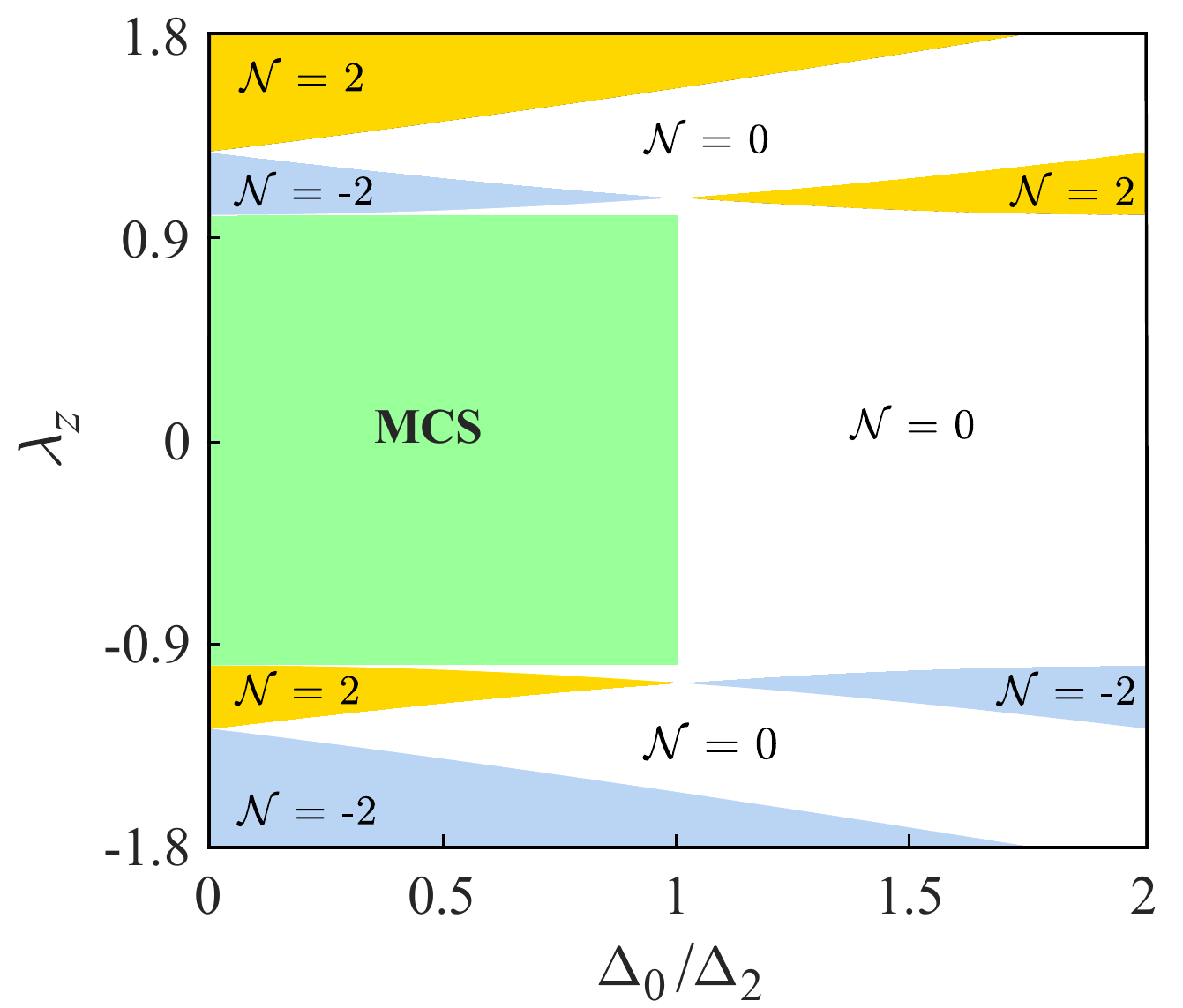}
	\caption{Topological phase diagram of a magnetic TI/$s_{\pm}$ high-temperature superconductor hybrid system in the $(\lambda_z, \Delta_{0}/\Delta_{2})$ plane, which reveals three distinct phases: (i) Second-order topological superconductor with MCSs, (ii) Chiral MZMs characterized by a finite Chern number $\mathcal{N}$, and (iii) Topologically trivial states $\mathcal{N} = 0$ with zero chiral MZMs and MCSs. Recall that $\Delta_{x} = \Delta_{y} = \Delta_{2}$, as shown in Eq.~(30). The parameters are chosen as $m_0 = -1$, $\lambda_{\textrm{so}}=1$, $m_x=-m_y=1$, $\mu = 0$, and $\Delta_{2} = 0.4$. }\label{fig7}
\end{figure}

\section{Discussion and Conclusion}
For experimental realizations,  we require a magnetic TI in proximity to a high-temperature superconductor. For the magnetic TI, we can consider the recently experimentally discovered TIs of 2D transition metal dichalcogenides (e.g., monolayer $\textrm{WTe}_2$ \cite{Wu76, Qian1256815}), or IV–VI semiconductors (e.g., monolayer PbS \cite{ZhangPbS2017, acs.nanolett.5b00308}), which coat a ferromagnetic insulator. The high-temperature superconductors could be the cuprate-based \cite{Zareapour2012, PhysRevB.91.235143, Wang2013} or iron-based  \cite{arXiv:1706.06074, Zhangeaan4596} materials, where topological superconductivity has been experimentally reported. It is thus quite attractive to study second-order TSC and possibly observe MCSs in these systems by considering their hybrids. Moreover, the magnetic exchange interaction in magnetic TI is usually highly tunable by external fields, and thus it is also interesting to study how the exchange interaction influences the features of second-order topological superconductivity.

In conclusion, we investigate the hybrid structure of a magnetic TI and a high-temperature superconductor, which exhibits second-order topological superconductivity. Both $d$-wave and $s_{\pm}$-wave superconducting pairing related to high-temperature superconductors are discussed. The hybrid systems are in the topologically trivial regime, but still support MBSs at each corner of a square sample. Because the hybrid systems preserve mirror-reflection symmetry, there are two MBSs at each corner in the whole regime of second-order topological phase studied here. We derive their corresponding topological phase diagrams, which emphasize the role of magnetic exchange interactions and pairing amplitudes. An intuitive edge argument shows that the corner states result from the opposite Dirac mass of two adjacent edges due to pairing symmetry. In the future, it would be interesting to look for experimental realizations of second-order TSCs, and study the possibility of non-Abelian braiding of MCSs in a 2D system.

\begin{figure}[hb]
	\centering
	\includegraphics[width=8.4cm]{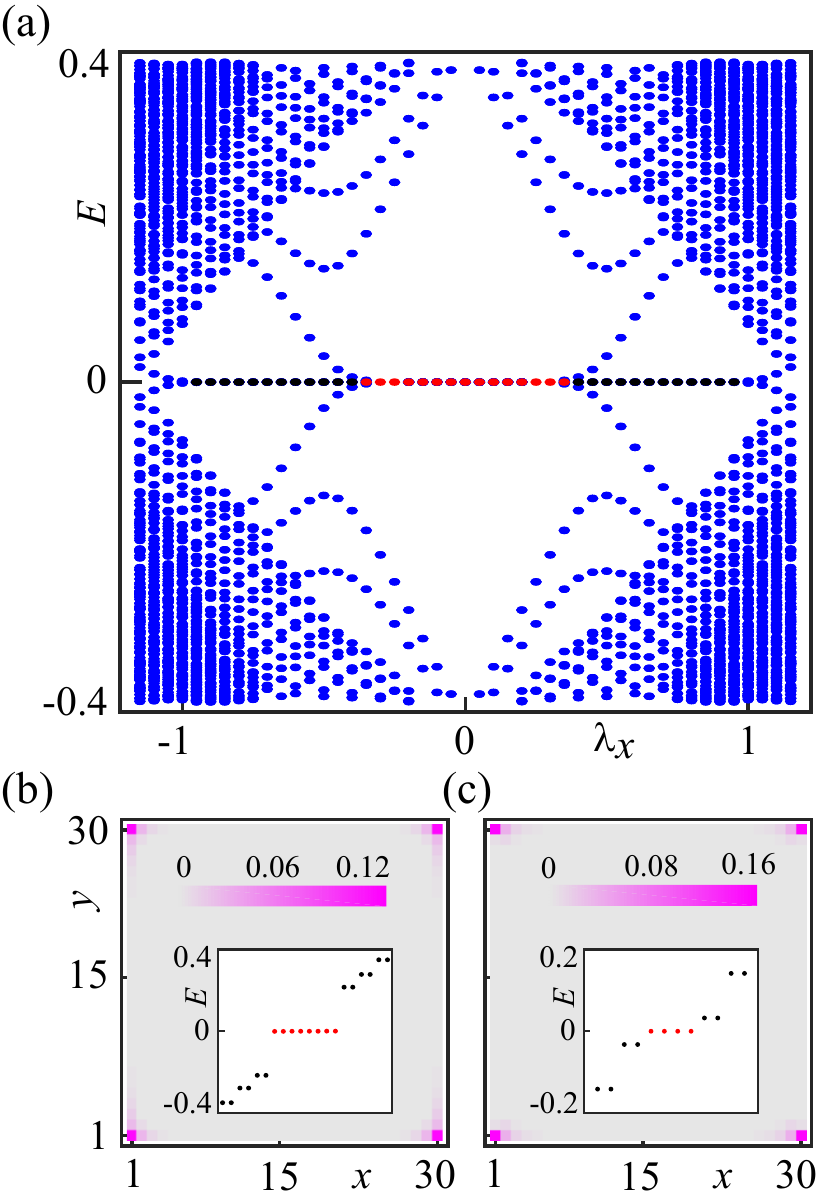}
	\caption{(a) BdG energy spectra of a $d_{x^2-y^2}$ wave pairing superconductor with open boundaries along the $x$ and $y$ directions as a function of $\lambda_x$. The red dots denote eight-fold degenerate MCSs, and the black dots represent the four-fold degenerate MCSs. The probability density distributions of mid-gap states for (b) $\lambda_x = 0.3$, and (c) $\lambda_x = 0.5$. The inset shows the eigenenergies with energies around zero. The parameters are chosen as $m_0 = -0.8$, $\lambda_{z} = 0.4$, $\lambda_{\textrm{so}}=1$, $m_x=m_y=1$, $\mu = 0$, and $\Delta_{1} = 0.5$.}
\end{figure}
\begin{figure}[hbt]
	\centering
	\includegraphics[width=8.4cm]{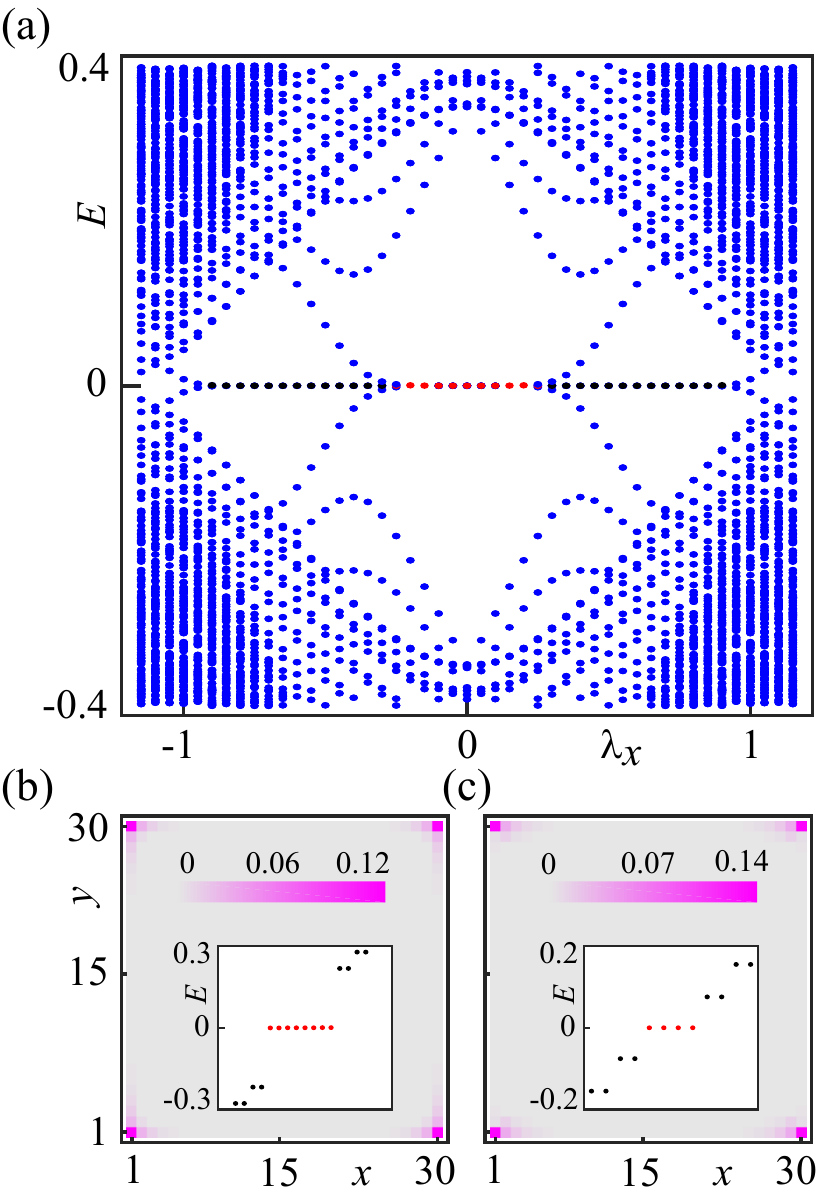}
	\caption{(a) BdG energy spectra of a $s_{\pm}$ wave pairing superconductor with open boundaries along the $x$ and $y$ directions as a function of $\lambda_x$. The red dots denote eight-fold degenerate MCSs, and the black dots represent the four-fold degenerate MCSs. The probability density distributions of mid-gap states for (b) $\lambda_x = 0.2$, and (c) $\lambda_x = 0.5$. The inset shows the eigenenergies with energies around zero. The parameters are chosen as $m_0 = -0.8$, $\lambda_{z} = 0.4$, $\lambda_{\textrm{so}}=1$, $m_x=-m_y=1$, $\mu = 0$, $\Delta_{0} = 0$, and $\Delta_{2} = 0.4$.}
\end{figure}

\begin{acknowledgments}
T.L. acknowledges support from a JSPS Postdoctoral Fellowship (P18023). F.N. is supported in part by the: MURI Center for Dynamic Magneto-Optics via the Air Force Office of Scientific Research (AFOSR) (FA9550-14-1-0040), Army Research Office (ARO) (Grant No. W911NF-18-1-0358), Asian Office of Aerospace Research and Development (AOARD) (Grant No. FA2386-18-1-4045), Japan Science and Technology Agency (JST) (the ImPACT program and CREST Grant No. JPMJCR1676), Japan Society for the Promotion of Science (JSPS) (JSPS-RFBR Grant No. 17-52-50023, and JSPS-FWO Grant No. VS.059.18N), RIKEN-AIST Challenge Research Fund, and the John Templeton Foundation.
\end{acknowledgments}

\appendix*

\section{Broken mirror-reflection symmetry}

The Hamiltonian $H_{\textrm{BdG}}(\mathbf{k})$ in Eq.~(5) respects the mirror-reflection symmetry $ \mathcal{M}_z $. Therefore, there are two MCSs at each corner in the whole regime of second-order topological phase considered here.  When this mirror-reflection symmetry is broken, a single MCS at each corner may be achieved for both $d$-wave and $s_{\pm}$-wave superconducting hybrid systems.

Let us now break the mirror-reflection symmetry by adding the term $\lambda_x s_x$ to Eq.~(5), so the Hamiltonian becomes
\begin{align}
\bar{H}_{\textrm{BdG}}(\mathbf{k}) & =   m(k) \sigma_{z} \tau_z + \lambda_{\textrm{so}} \left[\sin(k_x) s_x + \sin(k_y) s_y \right] \sigma_{x} \tau_z \nonumber \\
& ~~~  + \lambda_{z} s_z - \mu \tau_z + \Delta(k) \tau_x + \lambda_x s_x.
\end{align}

In the presence of $\lambda_x s_x$ term, the mirror-reflection symmetry is broken. Figure 8(a) shows the BdG energy spectrum of a $d_{x^2-y^2}$ wave pairing superconducting hybrid system with open boundaries along the $x$ and $y$ directions as a function of $\lambda_x$. The eight-fold degenerate MCSs exist [see red dots in Fig.~8(a) and probability density distributions in Fig.~8(c)] when the $\lambda_x$ is small, while there are only four-fold degenerate MCSs [see black dots in Fig.~8(b) and probability density distributions Fig.~8(c)] as $\lambda_x$ increases, where the second-order topological phase transition occurs. Therefore, a single MCS appears when the mirror-reflection is broken with an appropriate magnitude of $\lambda_x$

As in the case for a $d$-wave superconductor, for hybrid system with a $s_{\pm}$ wave pairing superconducting hybrid system, a single MCS at each corner can exist when the mirror-reflection symmetry is broken in the presence of the $\lambda_x s_x$ term with an appropriate magnitude (see Fig.~9). The effect of the breakdown of mirror-reflection symmetry on the MCSs in both $d$-wave and $s_{\pm}$-wave superconducting hybrid systems can be interpreted by considering the effective edge Hamiltonians derived from the Hamiltonian $ \bar{H}_{\textrm{BdG}}(\mathbf{k}) $ in Eq.~(A.1), as treated based on the block-diagonal Hamiltonian [see Eqs.~(18)--(29) and Eqs.~(31)--(38)].

\end{CJK*}

%\bibliography{MCS}
%merlin.mbs apsrev4-1.bst 2010-07-25 4.21a (PWD, AO, DPC) hacked
%Control: key (0)
%Control: author (0) dotless jnrlst
%Control: editor formatted (1) identically to author
%Control: production of article title (0) allowed
%Control: page (1) range
%Control: year (0) verbatim
%Control: production of eprint (0) enabled
%

\end{document}